\documentclass[latin1,final]{aa}
\usepackage{amsmath}
\usepackage{amssymb}
\usepackage{txfonts}
\usepackage{alltt}
\usepackage{inputenc}
\usepackage[T1]{fontenc}
\usepackage{natbib}

\usepackage{graphicx}
\usepackage[pdftex,colorlinks=false,pdfstartview=Fit,urlcolor=pdfwww,
  bookmarks=true,bookmarksnumbered=true]{hyperref}

\usepackage{url}
\usepackage{framed}

\bibpunct{(}{)}{;}{a}{}{,} 

\newcommand{\deleted}[1]{{\bf\ (DELETED TEXT)}}

\newcommand{\fnurl}[1]{\footnote{\url{#1}}}
\newcommand{\compl}[1]{\ensuremath{\mathcal{O}(#1)}}

\newcommand{\pnull}{\phantom{0}}

\newcommand{\Healpix}{HEALPix}

\newcommand{\ivco}[1]{\ensuremath{\left[{#1}\right[}}
\newcommand{\roundup}[1]{\ensuremath{\left\lceil{#1}\right\rceil}}
\newcommand{\rounddown}[1]{\ensuremath{\left\lfloor{#1}\right\rfloor}}
\newcommand{\vpla}{\vphantom{\LARGE A}}

\begin{document}

\title {Efficient data structures for masks on 2D grids}
\author {
M.~Reinecke\inst{\ref{mpa}}\and
E.~Hivon\inst{\ref{iap}}
}
\institute {Max-Planck-Institut f\"ur Astrophysik, Karl-Schwarzschild-Str.~1, 85741 Garching, Germany\label{mpa}\\ 
\email{martin@mpa-garching.mpg.de}\and
Sorbonne Universit\'es, UPMC Univ Paris 6 et CNRS (UMR 7095), 
Institut d'Astrophysique de Paris, 98 bis bd Arago, 75014 Paris, France\label{iap}\\
\email{hivon@iap.fr}
}

\date{Received May 18, 2015 / Accepted July 8, 2015}

\abstract {
This article discusses various methods of representing and manipulating arbitrary coverage information in two dimensions, with a focus on space- and time-efficiency when processing such coverages, storing them on disk, and transmitting them between computers.
While these considerations were originally motivated by the specific tasks of representing sky coverage and cross-matching catalogues of astronomical surveys, they can be profitably applied in many other situations as well.}

\maketitle

\section {Introduction}
\label{sect:intro}
\begin{figure}
\centerline{\includegraphics[width=0.8\columnwidth]{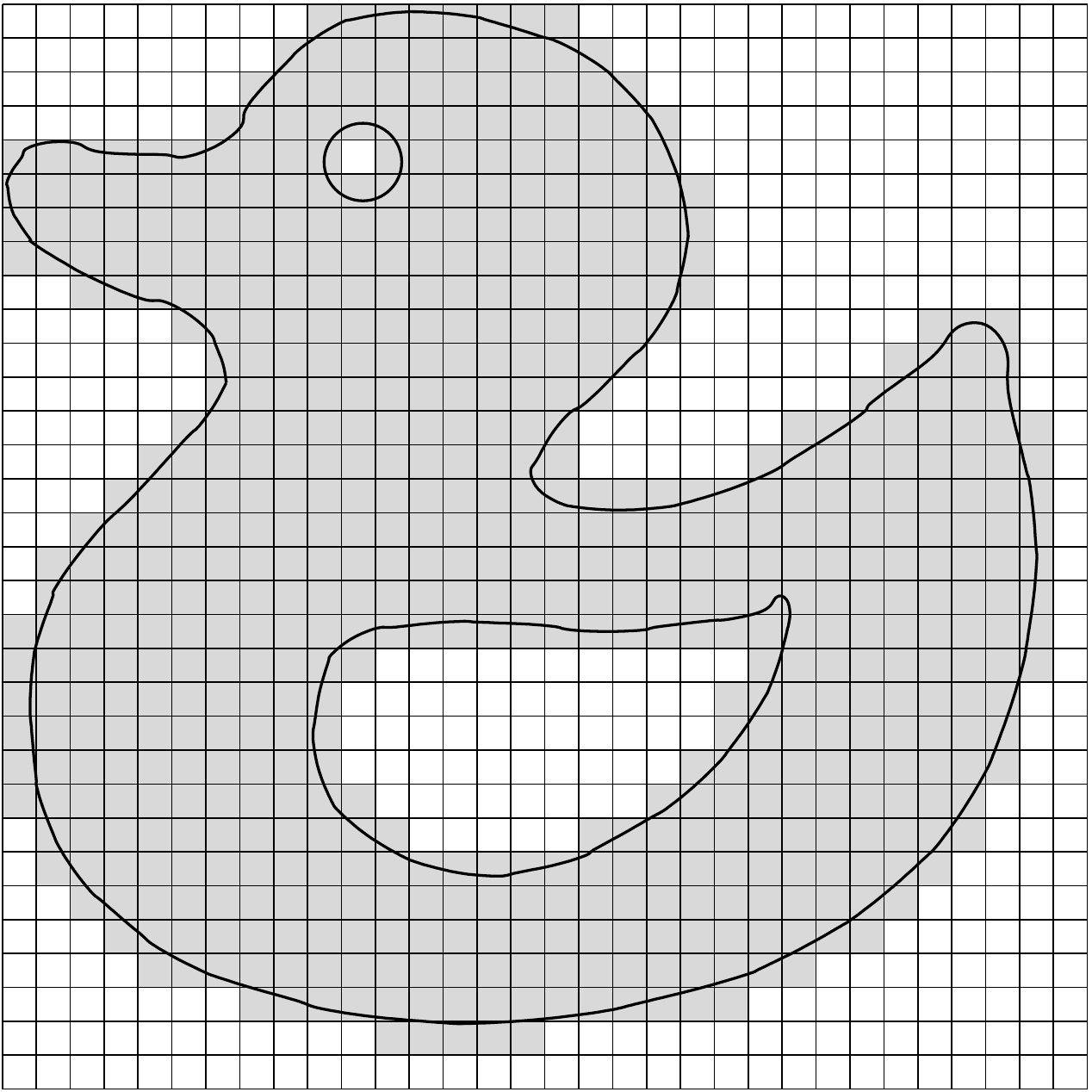}}
\caption{2D shape and its approximate discrete representation as a set of pixels.}
\label{figs:demo}
\end{figure}

This paper focuses on finding useful data representations for describing a subset of pixels on a 2D data grid (for an example see Fig.~\ref{figs:demo}). The boundary of this subset is sharp, which means that pixels can take one of exactly two values: 0 (outside the set) or 1 (inside).
The shape formed by the pixels can be arbitrary (it may, e.g., be disconnected and/or contain holes), but when judging the merits and drawbacks of the individual representations, it will be assumed that in typical scenarios, the shapes will not be pathological (e.g.\ they will not consist of clouds of isolated pixels or have fractally convoluted boundaries).

The notion of ``usefulness'' of course requires a usage scenario.
For this work, the concrete motivation was to improve support for astronomical databases with typical tasks, such as catalogue cross-matching (finding the overlap between two, potentially very complicated, shapes on the celestial sphere) and cone searches (i.e.\ obtaining all database objects lying within a certain shape on the sphere). It was inspired by the IVOA standard for multi-order coverage objects \citep{boch-etal-2014, fernique-etal-2015} and tries to evaluate potential shortcomings, as well as to suggest improvements.

The choice of describing shapes in an approximate way as a set of pixels on a grid rather than by analytical geometrical expressions was driven by efficiency concerns. Performing database queries using complicated analytical shapes is a difficult and time-consuming task, and trading a small, tunable number of false positives in the query result for much better run time is beneficial in many real-world situations.

The remainder of this paper is structured as follows.
Section \ref{sect:description} gives an in-depth description of the problem we set out to solve.
The subsequent sections discuss advantages and drawbacks of various approaches to the individual sub-tasks and justify our decision for a particular subset of them. Section \ref{sect:numbering} evaluates different ways to enumerate pixels on the grid, Section \ref{sect:uniquenumbers} introduces ways to incorporate resolution information into pixel numbers, Section \ref{sect:inmemory} discusses structure layouts for storage in main memory, and Section \ref{sect:compact} introduces ones for long-term storage and transmission.
A rough outline for an algorithm that generates approximate shape representations from analytical descriptions is given in Section \ref{sect:algorithm}.
Section \ref{sect:schemes} lists popular tesselations of the sphere and discusses whether our scheme can be applied to them, as well as explaining our choice of \Healpix\ for implementation and testing, which in turn is the subject of Section \ref{sect:tests}.
Finally, we present conclusions in Section \ref{sect:summary}.

\section{Detailed problem description}
\label{sect:description}
The task sketched above appears fairly straightforward at first glance, but it is certainly worthwhile to quickly present the many different aspects that need to be considered when attempting a solution.

\subsection{Requirements}
The introduction already hinted at two desirable features of the final data structure:
\begin{enumerate}
\item Boolean operations (merging, intersection, complement, tests for overlapping and containment) on shapes should be fast.
\item The choice of pixel numbering should provide good locality; that is,\ typical shapes should be described by relatively few ranges of consecutive pixel numbers. This property speeds up database queries related to the shape.
\end{enumerate}
An additional obvious design goal is that the data structure should be reasonably compact.
To be more precise:
\begin{enumerate}
\setcounter{enumi}{2}
\item The memory requirement of the data structure should grow, with increasing grid resolution, at a lower rate than the number of pixels covered by the shape. A realistic goal seems to be a growth rate proportional to the number of pixels crossed by the shape's boundary.
\end{enumerate}
For ``simple'' shapes like circles and squares, this implies a structure size proportional to the square root of the number of pixels in the shape. At the other extremes (highly convoluted or fragmented shapes), the growth rate will be higher, and it approaches that of the total number of pixels again, but in such a scenario, there is little hope for optimisation either.

In many situations the desired shape will be given by analytical expressions; for example, in the case of a simple cone search, this would be a circle. In astronomy, more complicated analytical shapes are often expressed using the STC grammar \citep{rots-2007} or the MANGLE tools \citep{swanson-etal-2008}.
Typically the boundary of such shapes does not coincide with the pixel boundaries of the chosen grid,
which leads to unavoidable inaccuracy of the discretised pixel-based representation, to a degree that depends on the grid's chosen resolution.

Since the derivation of a discretised representation from an analytic shape tends to be an expensive operation, and since it is quite common that the discretised representation is subsequently needed at several different resolutions, we require further that
\begin{enumerate}
\setcounter{enumi}{3}
\item The data structure must allow quick resolution changes (by factors of $2^n$ in both directions).
\end{enumerate}
For instance, going from a representation of a shape on a 64x64 grid to one on a 32x32 grid at half resolution and vice versa must be a very efficient operation.
The direction towards higher resolutions can obviously be performed without any loss of information, while the other direction leads to a coarsening of the discretised shape and additionally raises the question of how to treat partially covered pixels.
We require that coarsening operations must allow the user to specify whether such pixels should be kept as part of the coarser shape or discarded.

\begin{figure}
\centerline{\includegraphics[width=0.47\columnwidth]{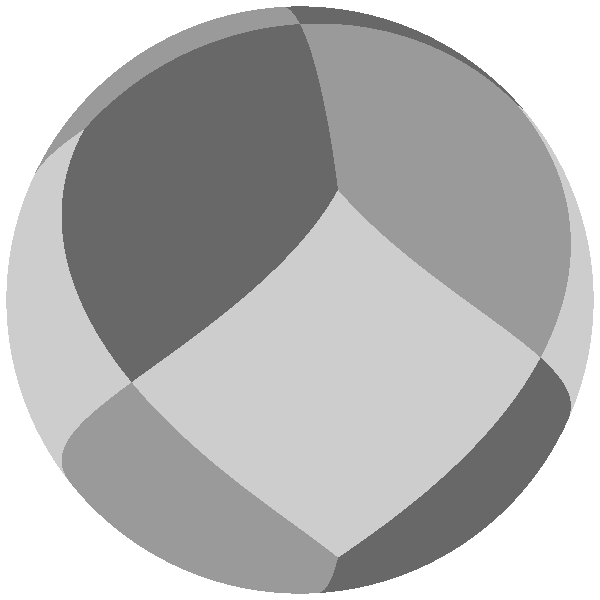}\hfill\includegraphics[width=0.47\columnwidth]{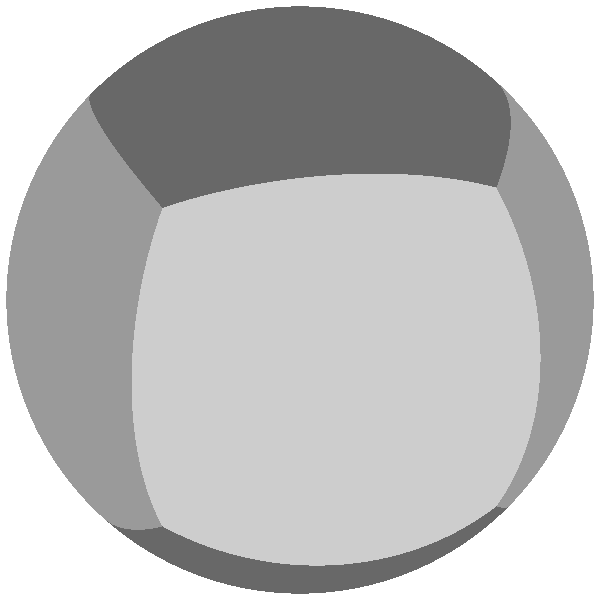}}
\caption{Subdivision of the sphere's surface into quadrangular patches by \Healpix\ (left) and QuadCube (right).}\label{figs:hp_quad}
\end{figure}

\subsection{Grid properties}
So far, only a single 2D grid patch with arbitrary dimensions in both spatial directions has been considered.
For practical purposes, this paper will focus on a slightly modified scenario.
On the one hand, we consider not only one patch, but instead a set of $N_0$ patches (or \textit{base pixels}) of the same size, on which the shape representation can reside.
On the other, we constrain all patches to the same dimensions of $2^o$ pixels in each spatial direction, where $o$ (for ``order'') is the single parameter determining the overall grid resolution. This constraint of patch resolutions to powers of 2 harmonises well with requirement 4 and has been adopted in practice by many pixelisation schemes.

As a practical example, these modified grid properties allow a very intuitive description of the \Healpix\ grid \citep{gorski-etal-2005} with its $N_0=12$ base pixels and its resolution parameter $N_\text{side}$, which is equivalent to $2^o$, or  QuadCube \citep{white-stemwedel-1992}, with $N_0=6$ base pixels and resolution described by the quantity $p:=o+1$ (see Fig.~\ref{figs:hp_quad}).
Constraining grid resolutions to powers of 2 may seem more draconian than it really is; if necessary, grids with other dimensions can be emulated by extending their dimensions to the next applicable power of 2 and by considering all pixels in the surplus area as unset.

\subsection{Accuracy of the shape representation}
As already mentioned above, discretising an analytical shape on a grid with finite resolution will (in most cases) introduce inaccuracies.
Using a simple illustration like Fig.~\ref{figs:demo}, it is straightforward to derive a rough estimate for the fractional error of the discrete representation:
\begin{equation}
\frac{A_\text{excess}}{A_\text{shape}} \propto \frac{L_\text{boundary}}{A_\text{shape}}d_\text{pix}\text{,}
\end{equation}
where $A_\text{excess}$ is the excess area of the discrete representation, $A_\text{shape}$ and $L_\text{boundary}$ are the original shape's area and boundary length, and $d_\text{pix}$ is the linear dimension of a grid pixel.
Of course this estimate is only applicable on ``well-behaved'' grids where pixels have almost equal sizes and are not very elongated; otherwise, the linear pixel dimension would not be a well-defined quantity.

The estimate illustrates that the fractional error scales linearly with the grid resolution;
thus, for a specific desired error tolerance, it is possible to choose an appropriate pixel size whenever computing the approximate discrete representation of an analytical shape.
Since $d_\text{pix}$ cannot become larger than the linear dimension of a base pixel, the relation above naturally holds only on smaller scales.
Given the typically small numbers of base pixels in spherical pixelisation schemes -- 12 for \Healpix, 6 for QuadCube, 8 for HTM \citep{szalay-etal-2007}, 12 for basic IGLOO \citep{crittenden-turok-1998} -- this is not a real problem in most cases.
However, for pixelisation schemes with very many base pixels (e.g.\ some of the more involved IGLOO tilings that can have $N_0>10~000$), the coarsest possible discretisation of a shape may be finer (and therefore more memory-consuming) than required by the application. 

\section {Evaluation of pixel numbering schemes}
\label{sect:numbering}

Starting from the goals established in Section \ref{sect:description}, the first task is to assign indices to the individual pixels in the 2D grid, preferably in a way that aids compact representation of shapes.
Assuming $N_0$ base pixels and a resolution order $o$, there are $n_{\text{pix,}o}:=N_0 2^{2o}$ pixels in total on the grid, and giving them numbers from the interval \ivco{0;N_0 2^{2o}} seems a reasonable choice.
Analogously, it is most likely a good idea to group pixels belonging to the same patch together, i.e.\ associating indices \ivco{0;2^{2o}} with the first patch, indices \ivco{2^{2o};2\times2^{2o}} with the second patch, etc.
Obviously this choice of pixel numbers only identifies a pixel within the grid at a specified order; for an approach to encoding both order and location in the grid in a single number, see Sect.~\ref{sect:uniquenumbers}.

\subsection{Linear ordering}
\begin{figure}
\centerline{\includegraphics[width=0.8\columnwidth]{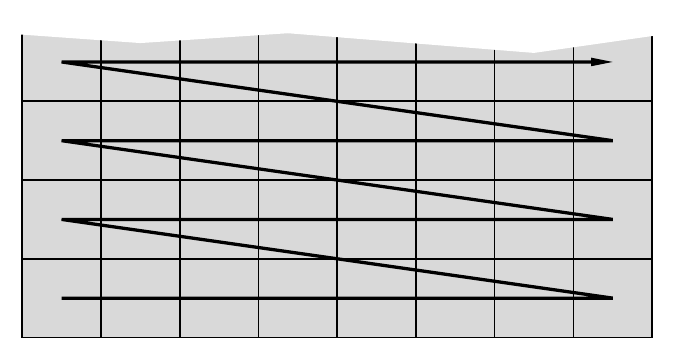}}
\caption{Traditional linear numbering scheme.}\label{figs:linear}
\end{figure}
 Several options are available, however, for the indexing scheme for the pixels within a patch.
A very well-known and intuitive choice would be a linear ordering that counts first
all pixels in the first row of the patch, then those in the second row, etc.\ (see Fig.~\ref{figs:linear}).
While very easy to implement, this arrangement is sub-optimal for the purpose of compact shape representation:
the traversal of a full spatial dimension for every row implies bad locality properties (conflicting with requirement 2). 
Another, less severe drawback is the nontriviality of resolution change operations (requirement 4): while not exactly daunting, grid refinement and coarsening are more complicated than in other, more promising numbering schemes.

\subsection{Hierarchical schemes}
\label{sect:hierarch_numbering}
Given the above-mentioned shortcomings of the most straightforward approach to pixel numbering, it seems worthwhile to specifically investigate schemes that have some sort of resolution hierarchy built in.
One family of such schemes can be constructed using the simple recursive rule that a pixel with the number $p$ at resolution order $o$ must coincide with the four pixels \ivco{4p;4(p+1)} at order $o+1$.

For every numbering scheme that follows this rule, requirement 4 is trivially fulfilled: $p$'s parent pixel at order $o_2<o$ has the number \rounddown{p/2^{2(o-o_2)}}, and at higher orders $o_3>o$, it covers the range of pixels \ivco{p\times2^{2(o_3-o)};(p+1)\times 2^{2(o_3-o)}}. All necessary conversions between these numbers can be performed using extremely quick bit-shift operations.

While reducing the possible numbering choices, the above constraint does not specify a unique ordering in itself: there is still the choice of how exactly the available indices are assigned to the four sub-pixels when going from order $o$ to $o+1$.
Since this choice can be made differently for each single pixel and at each order, the number of theoretically available numberings is huge, but fortunately, only a handful of these have attractive geometrical and algorithmic properties.

\begin{figure}
\centerline{\includegraphics[width=0.9\columnwidth]{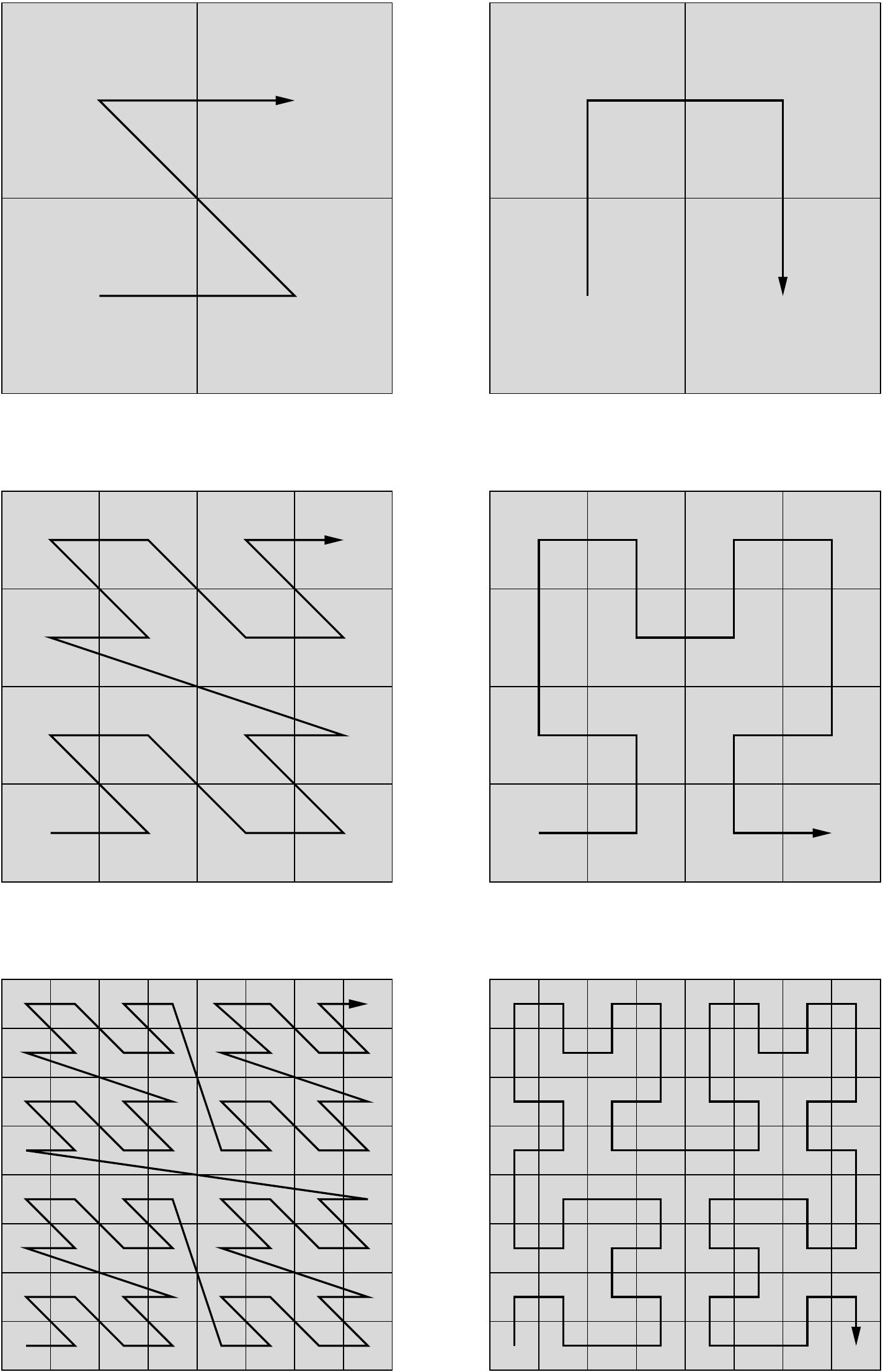}}
\caption{Three levels of refinement for the Z curve (left) and Peano-Hilbert curve (right).}\label{figs:curves}
\end{figure}

Figure~\ref{figs:curves} shows two well-studied variants known as the Morton (or Z) and Peano-Hilbert curves.
For both subdivision strategies, the recursive nature and self-similarity of the pixel ordering is clearly visible.
The figure also demonstrates intuitively that the Peano-Hilbert curve tends to have better locality than the Z curve, because the geometrical distance between Peano-Hilbert pixels $n$ and $n+1$ is always the shortest possible one (i.e.\ the linear dimension of a single pixel), whereas the Z curve  exhibits fairly large spatial jumps between pixels of neighbouring indices.

Unfortunately, the improved locality properties of Peano-Hilbert ordering come at a price: converting the two-dimensional location ($x_i$, $y_i$) of a pixel in the grid at order $n$ to its corresponding Peano-Hilbert index is an iterative process with $n$ individual, nontrivial steps, and the same holds for the inverse operation \citep{lam-shapiro-1994}. The same tasks for the Z curve, however, can be achieved by comparatively simple bit manipulations and, on modern CPUs, even by using specialised machine instructions.\footnote{PDEP and PEXT, see e.g. \url{http://en.wikipedia.org/wiki/Bit_Manipulation_Instruction_Sets}. Unfortunately, compiler support is not yet wide-spread.}

Whether efficient index calculations should be preferred over locality (hence compactness of representation) or vice versa depends on the task at hand. It should be noted, however, that, among all algorithms discussed in this paper, only the initial generation of a pixelised shape representation requires any conversion of pixel indices. All other operations (such as resolution changes, Boolean operations on one or multiple shapes, or conversion to long-term storage format) are completely oblivious to the the chosen numbering scheme and will only benefit from good locality, but not suffer from slow index computations. We expect that once created, a given shape will usually be processed in queries many times, so that any slowness during creation will be more than amortised by the speedups of a compact representation; therefore, Peano-Hilbert should be the preferred ordering.

Another possible choice for hierarchical ordering, based on Gray codes\footnote{A method of binary representation of integer numbers with (probably accidental) hierarchical properties; see \url{https://en.wikipedia.org/?title=Gray_code} and \cite{moon-etal-2001}.}, should be noted for completeness. Its locality lies between those of Z and Peano-Hilbert curves (see, e.g., \citealt{moon-etal-2001} for benchmarks), and its index computations are about as complex as those of Peano-Hilbert. Consequently we will not investigate this option further.

\section {Unique multi-order indices}
\label{sect:uniquenumbers}
Section~\ref{sect:numbering} discussed various choices for numbering pixels within the grid at a given resolution order. The resulting numbers do not carry any information about the order of the grid they refer to, so this quantity must be known from other sources. In some circumstances it is convenient to encode this resolution information together with the pixel index in a single number, and this section discusses economical ways to achieve this goal.

The simplest approach from an implementation standpoint would be to reserve a few bits of the pixel number for directly storing the order. This allows conversions from (order + pixel index) to unique pixel number and vice versa in constant time, using only simple bit shift and bit masking operations. It has, however, the drawback that the number of bits required for storing the order is \roundup{\log_2 o_\text{max}}, which is quite wasteful compared to alternative approaches. Potentially worse, it requires the agreement on a common $o_\text{max}$ over all involved applications.

Alternatively, it is possible to start numbering the pixel indices for order $n$ from an offset $O_n$ instead of 0, with the no-overlap constraint that $O_{n+1}$ is greater than the highest pixel number at order $n$:
\begin{equation}
O_{n+1}\ge O_n+n_{\text{pix},n}\text{.}
\end{equation}

Pixel numbers can be minimised by choosing $O_0:=0$ and the equality $O_{n+1}:= O_n+n_{\text{pix},n}$.
Using this dense packing, and because in two dimensions $n_{\text{pix},n}/n_{\text{pix},n-1}=4$, the maximum unique pixel index at order $n$ is never larger than $4n_{\text{pix},n}/3$. This implies that the necessary bit length for the unique pixel index is at most one bit higher than for the standard pixel index at any resolution, which compares favourably to the approach described above. As an aside, this growth of at most one bit is also true for 1D and all higher-dimensional scenarios.

While conversion from (order, pixel number) pairs to unique indices is still a \compl{1} operation in this scheme, the inverse operation becomes more expensive in the general case, because determining the order corresponding to a given unique pixel index can only be done via interval bisection in \compl{\log_2\, o_\text{max}} steps, which is undesirable.

By giving up the dense packing and allowing for some unused pixel indices between the blocks for the various resolution orders, this drawback can be overcome. If the $O_n$ are constrained to be powers of 2, determining the order of a given unique pixel number $i$ becomes equivalent to computing \rounddown{\log_2 i}, which can be done in constant time by specific machine instructions on most current CPUs.
The general formula for the smallest possible $O_n$ that is a power of 2 for a given $N_0$ reads as
\begin{equation}
O_n=2^{2n}\times [2^{\roundup{\log_2(4N_0 /3)}} - N_0]\text{.}
\end{equation}

As a practical example, for \Healpix\ with $N_0 = 12$, one obtains $O_n=2^{2n+2}$, which means that the 12 pixels at order 0 have the unique indices 4 to 15, those at order 1 have 16 to 63, etc. Interestingly, the number of bits required to represent a \Healpix\ pixel at any order does not increase when switching to unique indices.

It is also worth mentioning that in the above example and, generally, in all cases where $O_n$ is an integer  multiple of $2^{2n}$, the convenient hierarchy property described in Section \ref{sect:hierarch_numbering} holds for unique pixel numbers as well.

\section {In-memory representation}
\label{sect:inmemory}

In this section we discuss alternative data structures describing a shape on a chosen 2D grid, assuming one of the hierarchical pixel ordering schemes presented in Section \ref{sect:hierarch_numbering}.
The goal is to determine the structure most suitable for Boolean operations and resolution changes. Another, more specialised, format for storage and transmission is presented in Section \ref{sect:compact}.

The naive choice of simply storing a list of all pixels at a given resolution that lie within the shape can be discarded immediately as impractical. The number of entries in such a list would always scale with \compl{2^{2o}} in direct contradiction to requirement 3 and would very quickly become unmanageable in typical scenarios. As an example, a shape covering ten percent of the sphere and stored at a resolution of 1 arcsec would contain roughly $5\cdot 10^{10}$~pixels, which does not fit into most computers' main memory.

\subsection{Multi-order list}
\label{sect:mol}
One way to avoid the prohibitive size of a full pixel list at the highest resolution was suggested in the MOC standard document \citep{boch-etal-2014}, and it exploits the hierarchical property of the pixel numbering scheme:
instead of having a single list of pixels at the highest resolution $o_\text{max}$, a sorted list is kept for every resolution order $0\le o \le o_\text{max}$.
Starting from $o=0$, all pixels at this resolution that are completely covered by the shape are entered into the list for this order. Then the procedure is repeated for the remainder of the shape at the next higher resolution, until the desired resolution is reached.

This procedure yields a very compact representation of the shape (called multi-order list, or MOL, below), which certainly fulfils requirement 3. Also, this structure facilitates resolution changes. Resolution upgrade is an empty operation, and coarsening simply consists of removing the pixel lists above the desired resolution, if partially covered pixels are ignored; if they should be kept, the pixel lists have to be updated recursively from higher to lower resolutions.

Unfortunately, multi-order lists are not well suited to Boolean operations. To our knowledge, no algorithm exists to compute, for example, the union of two shapes in \compl{n_1+n_2} time, where $n_1$ and $n_2$ are the respective total lengths of both shapes' multi-order lists. Thus, the MOL representation does not satisfy requirement 1.

It should be noted that, by construction, the total size of a multi-order list for a given shape does not depend on the particular choice of hierarchical numbering scheme; in other words, the MOL representation does not benefit from the improved locality of the Peano-Hilbert curve compared to that of the Z curve.

\subsection{Range set}
\label{sect:rs}
Since storing lists of pixels at various resolutions complicates Boolean operations, a preferable method is to simply store a sorted collection of \textit{\emph{ranges}} of pixels (characterised by their beginning and end) at the highest resolution. We refer to this structure as a \textit{\emph{range set}} (RS).

There are different ways to represent a range of numbers: One could use the first number plus the range's length or the first and last numbers of the range, for instance. Mostly for reasons of symmetry, we are adopting a representation that is commonly used by programmers by storing the first number of the range and the first number \textit{\emph{after}} the range. While this choice may seem unintuitive, it is technically preferable in many small ways over the other representations.\footnote{As an example, it gives symmetry to ``on'' and ``off'' ranges, making inversion operations trivial to implement. Also, this approach of describing a range set leads to a strictly increasing sequence of numbers, which is convenient for compression (see Sect.~\ref{sect:compact}).}

Even though formally residing on a high-resolution grid alone, the range set representation still benefits substantially from the hierarchical  numbering scheme: each pixel at resolution order $o$ that is completely filled is represented by a single interval of pixel numbers at higher resolutions $o_2>o$, by construction. This allows a worst-case size estimate of the RS representation compared to MOL: assuming that no pixel ranges can be merged, every pixel in the MOL becomes an isolated pixel range in the corresponding RS. Since a range is characterised by two numbers instead of one, the RS representation is, at worst, twice as large as the MOL.
 
There are also situations where the RS is smaller than the MOL, however: assuming a shape that covers the entire grid except for a single pixel at order $n$, the MOL will consist of $N_0-1 + 3n$ pixels, whereas the RS will contain at most two ranges. Generally, the more regular a shape (i.e.\ the lower its ratio of boundary length to surface), the smaller its RS representation will be.

In practice both representations tend to have a similar size for compact shapes, with the RS typically being slightly smaller than the MOL. For very convoluted or fragmented shapes, the MOL representation has the advantage, and the size ratio approaches the limit of 2:1 in favour of the MOL. See Sect.~\ref{sect:mem_benchmarks} for real-world examples.

In contrast to multi-order lists, RSs do benefit from better locality properties of the underlying pixel numbering scheme, which means that\ a RS representation of a shape on a Peano-Hilbert-indexed grid has, on average, fewer and longer ranges than on a Z-curve-indexed grid. This reduction in the number of ranges is more pronounced for ``regular'' shapes and vanishes in the limit of complete fragmentation (see the third row of Fig.~\ref{figs:memory} for real-world examples). When the pixel numbering is used to index entities in a database, the improved compactness of the Peano-Hilbert representation improves query times.

Range sets behave very similarly to multi-order lists for resolution changes. Resolution increases are again empty operations, and decreases~-- whether inclusive or exclusive~-- require simple adjustments of range borders and potentially merging of ranges, which scales linearly with the number of ranges in the set.

The substantial advantage of RS over the MOL representation lies in the simplicity and efficiency of all Boolean operations. This includes union and intersection of two shapes, as well as finding the complement of a shape and testing whether a shape contains, overlaps with, or is equal to another. These operations can be carried out in at most \compl{n_1+n_2} steps, where $n_1$ and $n_2$ are the range counts in both involved shapes. For the subset and overlapping tests, alternative algorithms with the complexity \compl{n_\text{min} \log n_\text{max}} (with $n_\text{min}:=\text{min}(n_1\text{,}n_2)$ and $n_\text{max}:=\text{max}(n_1\text{,}n_2)$) may be used, which can be advantageous if both shapes have very different range counts. In most situations, this advantage should more than balance out the drawback of the potentially larger RS size compared to MOL.

In fact, the most efficient way to perform Boolean operations on MOL that the authors have discovered so far is to convert the input MOL to RSs, perform the desired operation, and convert the result back to MOL. (This is also the approach recommended by the MOC standard.) The conversions between both representations have a slightly higher complexity than the Boolean operation itself, since they involve sorting and merging of sorted pixel lists at all involved resolution levels.

\section {Long-term storage and transmission format}
\label{sect:compact}
Section~\ref{sect:inmemory} presented a data structure suitable for fast data processing. In this section the focus is shifted towards finding a representation of the 2D shape that does not directly support Boolean operations, but is even more compact, while still being efficiently convertible to the RS or MOL representation. This data structure could be used for space-efficient storage on disk or for quick transmission over a network. In other words, we can relax our computation-related requirements 1 and 4 and concentrate mainly on size-related requirement 3 in this section.

In this context it is very helpful to observe that both the range set representation and the multi-order list representation (assuming unique pixel indices, see Sect.~\ref{sect:uniquenumbers}) of a shape take the form of strictly monotonous sequences of non-negative integers. In addition, these integers tend to build~-- at least for non-pathological shapes~-- relatively compact clusters within the possible range of values.

As such, these sequences are perfectly suited for \textit{\emph{binary interpolative coding}} \citep{moffat-stuiver-2000}, a compression algorithm typically used for lookup tables in search engines. This method provides a fairly simple and quick means to convert between the sequence and a highly compressed bit stream containing equivalent information. The algorithm has the complexity \compl{n}\ in both directions for a sequence of $n$ elements and can be implemented and customised for the task at hand using no more than a few hundred lines of code. Compression factors typically range from 2 to 10, and the time required for conversion is roughly comparable to the conversion time between MOL and RS representations. For concrete performance measurements on a large set of test data see Sect.~\ref{sect:tests}.

\section {Generation from analytical shapes}
\label{sect:algorithm}
Creating sets of pixels that approximately represent an ideal geometric shape can be achieved in many different ways. We briefly present an approach that poses minimal requirements on both the description of the shape and the grid geometry, has low complexity, and is conceptually easy to implement.

In addition to the algorithms discussed so far, only one further ingredient is required: a function which, given a specific pixel and order, returns whether
\begin{itemize}
\item the pixel lies completely within the shape;
\item the pixel lies completely outside the shape;
\item the pixel centre lies inside the shape, but it cannot be decided whether the pixel lies on the boundary or is completely inside; or
\item the pixel centre lies outside the shape, but it cannot be decided whether the pixel lies on the boundary or is completely outside.
\end{itemize} 
To obtain the desired set of pixels, the following algorithm is executed for all $N_0$ base pixels:
\begin{itemize}
\item test the current pixel with the function given above;
\item if it is completely inside the shape, append it to the result pixel list;
\item (if it is completely outside, do nothing);
\item otherwise:
\begin{itemize}
  \item if the order $o$ of the current pixel is smaller than the maximum order $o_\text{max}$ for this query, call the algorithm recursively for its four sub-pixels.
  \item if $o=o_\text{max}$ and the centre is inside the shape, append the pixel to the result pixel list.
  \item if $o=o_\text{max}$ and the centre is outside the shape, append the pixel to the result pixel list only if the query is inclusive.
\end{itemize} 
\end{itemize} 
Here, an \textit{\emph{inclusive query}} means that all pixels potentially overlapping with the shape are included in the result, in contrast to standard queries, which simply return all pixels whose centres lie inside the shape. While standard queries give exact results, the results of inclusive queries may contain a small number of false positives, which are\ pixels that do not actually overlap with the shape. Given the simple inside/outside criterion described above, this is unavoidable.

This algorithm has the advantage of checking high-resolution pixels only near the shape's boundary. Far away from the boundary (whether inside or outside), the status of the checked pixels can already be determined at low resolutions, so that the recursion terminates early. For reasonably compact and non-convoluted shapes, this leads to a complexity proportional to the length of the resulting RS, which is very welcome.

The above algorithm has been used in almost exactly this form at the core of the \texttt{query\_disc} and \texttt{query\_polygon} routines of the \Healpix\ C++ library for several years and has proven very robust.

\section{Applicability to popular pixelisation schemes}
\label{sect:schemes}
The techniques developed above can, in principle, be applied to any hierarchical grid. The benefits, however, depend on the individual grid's properties. In the following, we present some of the most popular spherical pixelisations and quickly discuss the feasibility and possible limitations of applying our techniques to them.

\begin{description}
\item[ECP:] This long-known grid with pixel centres equidistant in latitude and longitude can be designed to be hierarchical. However, the grid cells are extremely elongated near the poles, and the pixel areas vary strongly, so this grid is not suited to our purpose.

\item[QuadCube \citep{white-stemwedel-1992}:] With fairly uniform pixels and a hierarchical structure, this pixelisation should be quite suitable.

\item[IGLOO \citep{crittenden-turok-1998}:] This grid is hierarchical by design and has reasonably uniform pixel sizes and compact pixel shapes. It should be a suitable basis for our shape representations, but owing to the specific refinement procedure near the poles, it might be hard to find traversals with good locality.

\item[\Healpix\ \citep{gorski-etal-2005}:] Designed to be hierarchical with exactly equal pixel areas and fairly uniform pixel shapes, this pixelisation is very well suited to our studies.

\item[GLESP \citep{doroshkevich-etal-2005}:] Its geometrical requirements mean that this grid is inherently non-hierarchical.

\item[HTM \citep{szalay-etal-2007}:] This grid is inherently hierarchical. That its pixels are triangular does not present a problem, since recursive refinement in factors of 4 also works in this case. In combination with fairly uniform pixel sizes and shapes, it is a suitable underlying grid for our shape-representation algorithms. The only real concern is the choice of subpixel numbering: since the subpixels in the triangle corners get indices 0-2, and the one in the centre gets index 3, the traversing curve has bad locality, resulting in unnecessarily large RSs.
This could be mitigated by using a modified counting scheme, however.

\end{description}

For the tests presented in this paper, we have adopted the \Healpix\ grid for the following reasons:
\begin{itemize}
\item The MOC standard already specifies \Healpix\ as underlying pixelisation, making results obtained for this grid especially relevant.
\item Our test data set (see next section) was already provided as \Healpix\ MOC objects; re-discretisation on a different grid would have been possible, but cumbersome, without providing additional value.
\item The existing software for \Healpix\ is extensive and easily accessible, and already contains a considerable part of the required functionality (e.g.\ the conversion routines between NEST and Peano-Hilbert indices).   
\item Given our involvement in the development of the \Healpix\ code, working with this software was the most efficient choice.
\end{itemize}

\section {Validation and performance tests}
\label{sect:tests}
The practical application of the algorithms and data structures presented above will be demonstrated in the framework of the \Healpix\ C++ library. This is convenient, since the so-called NESTED ordering scheme for \Healpix\ pixels is equivalent to Z-curve ordering, and since the C++ library also supports Peano-Hilbert indexing of pixels. This was added to allow the research presented by \cite{schaefer-2005}.

For our validations we made use of the very extensive collection of astronomical sky coverages available at \url{http://alasky.u-strasbg.fr/MocServer/MocQuery}, which at the time of download consisted of 14~633 data sets, stored as multi-order lists in FITS files.
These contain very small, low-resolution shapes as well as large, but compact, survey coverages and highly fragmented star catalogues at high resolutions. MOL sizes range from roughly ten to several million entries, and maximum resolution orders vary between 0 and 19 (corresponding to a linear pixel dimension of $\sim$0.5~arcsec).

\subsection {Validation}

To test the correctness of our implementation, we verified a series of identities, using the shapes in our data collection. In the following, $A$ and $B$ denote shapes represented as a RS in Z-order indexing.

The following tests involving single data sets were performed on all available shapes:
\begin{itemize}
\item FITS input/output:\\ $A\overset{?}{=} \text{fromFITS(toFITS(}A\text))$
\item conversion to/from MOL:\\ $A\overset{?}{=} \text{fromMOL(toMOL(}A\text))$
\item compression:\\ $A\overset{?}{=} \text{uncompress(compress(}A\text))$
\item MOL compression:\\ $A\overset{?}{=} \text{fromMol(uncompress(compress(toMOL(}A\text))))$
\item Peano-Hilbert conversion:\\ $A\overset{?}{=} \text{fromPeano(toPeano(}A\text))$
\item complement:\\ $A \cap \overline{A} \overset{?}{=} \emptyset\text{; }A \cup \overline{A} \overset{?}{=} \text{entire sphere}$
\end{itemize}
Furthermore, we verified the following identities for a subset of all possible shape pairs:
\begin{itemize}
\item $\left(A\cap \overline{B}\right)\cap B \overset{?}{=} \emptyset$
\item $A\cup B \overset{?}\supseteq A \text{; } A\cup B \overset{?}\supseteq B $
\item $A\cap B \overset{?}\subseteq A \text{; } A\cap B \overset{?}\subseteq B $
\item $A\wedge B \overset{?}= (A\cup B) \cap \overline{(A\cap B)} \overset{?}=  \left(A\cap \overline B\right) \cup \left(\overline A\cap B\right)$.
\end{itemize}
The subset was obtained by first sorting the available shapes according to their number of ranges and picking those at positions $p_i:=147i$, resulting in 100 shapes sampled fairly from all available complexities. All 10~000 shape pairs constructible from this subset were tested.

\subsection {CPU benchmarks}
\label{sect:cpu_benchmarks}

All tests were performed on a single core of an Intel Core i3-2120 CPU running at 3.3GHz, using \texttt{gcc}~4.7.3 as compiler.

\subsubsection{Conversions between different representations}
\label{sect:conversions}
\begin{table}
\begin{center}
\begin{tabular}{l|ccccc}
        \vpla       &   RS Z   & RS P & MOL & CRS Z & CRS P  \\ \hline
        \vpla RS Z  &   ---    & 489  & 100 &  176  &  656   \\
        \vpla RS P  &   472    & ---  & 391 &  648  &  167   \\
        \vpla MOL   & \pnull80 & 386  & --- &  256  &  553   \\
        \vpla CRS Z &   166    & 655  & 266 &  ---  &  822   \\
        \vpla CRS P &   632    & 160  & 551 &  828  &  ---   \\
\end{tabular}
\end{center}
\caption{Overview of average conversion times from the shape representation in the top row to the one in the first column. The abbreviations denote ``range set in Z ordering'', ``range set in Peano-Hilbert ordering'', ``multi-order list in Z ordering'', and the respective compressed variants of the range sets. All times are given in microseconds.}
\label{table:conversions}
\end{table}

Sections \ref{sect:inmemory} and \ref{sect:compact} discussed various data structures for representing a 2D shape with the conclusion that, depending on the concrete usage scenario and the compactness of the shape, no single one fits all needs perfectly. Consequently, conversions between different representations may occur frequently, and it is important that they can be carried out quickly.

Table \ref{table:conversions} lists the conversion times between several representations, averaged over all shapes in the test data set. Obviously, converting from a MOL to a RS in Z ordering and vice versa only takes around 100 microseconds on average, which is advantageous, because this kind of conversion is likely needed most often.

Conversions between compressed and uncompressed RSs of the same scheme are more expensive, but only by about a factor of two. This makes compressed range sets a very attractive choice for shape storage whenever memory is at a premium.

Finally, changes between the Z-curve and Peano-Hilbert pixel numbering schemes are fairly costly, so it is probably best to decide on an overall numbering scheme for each application, and to perform this sort of conversion only during data exchange with other external programs, if unavoidable.

\subsubsection{Cost of Boolean operations}
\begin{table}
\begin{center}
\begin{tabular}{l|l|c}
        \vpla XOR&$A \wedge B$       &  98.30  \\
        \vpla OR&$A\cup B$         &  72.36  \\
        \vpla ANDNOT&$A\cap \overline B$     &  34.93  \\
        \vpla AND&$A\cap B$        &  \pnull5.66  \\
        \vpla overlap&$A\cap B \overset{?}{=} \emptyset$   &  \pnull1.66  \\
        \vpla containment&$A\cup B \overset{?}{=}A$   &  \pnull0.19  \\
        \vpla complement & $\overline A$ & 43.80
\end{tabular}
\end{center}
\caption{CPU times for various unary and binary Boolean operations, averaged over all possible combinations of shapes in the available data set. Shapes are stored as range sets in Z ordering. All times are given in microseconds. }
\label{table:boolperf}
\end{table}
Table~\ref{table:boolperf} lists average execution times for the Boolean operations on the shapes that are supported by the library. Probably the most noteworthy fact is that all operations are faster than (or in the worst case have run times similar to) the conversions discussed in Sect.~\ref{sect:conversions}. This implies that storing the shapes using a non-RS representation and converting to RS whenever needed incurs a substantial slowdown and should therefore only be done when computer memory is scarce.

Amongst the binary operations, the exclusive-or operation consumes the most time. This is according to expectations, since XOR requires an examination of every range start- and endpoint for both involved RSs, while shortcuts are possible for AND, OR, and ANDNOT. Also, a RS constructed from two others via XOR tends to contain more ranges than ones constructed from the same sources using OR or AND.

Computing the union (OR) of two shapes is more expensive on average than intersection (AND) or subtraction (ANDNOT), which might be because RSs of unions are typically longer than those of the last two operations. That construction of the result requires a large amount of the algorithm's runtime. The pronounced difference between the average times for AND and ANDNOT is most likely caused by the nature of our input data set: most of the shapes only cover a very small part of the sphere, so intersections often result in empty or very small sets, whereas subtractions typically reproduce the first operand and consequently need more time to construct their output.

The checks whether one shape overlaps another or is completely contained within another require considerably less CPU time. This is unsurprising, since the test can exit early as soon as the first overlap (or the first non-containment) is found, and since both algorithms return only a single Boolean value and do not have to construct an output RS. Again, owing to the nature of the test data set, which contains many small and very often disjoint shapes, early exit is much more likely for the containment tests, thereby lowering the average CPU time for this case.

\subsection {Memory benchmarks}
\label{sect:mem_benchmarks}
\begin{figure*}[p]
\includegraphics[width=0.5\textwidth]{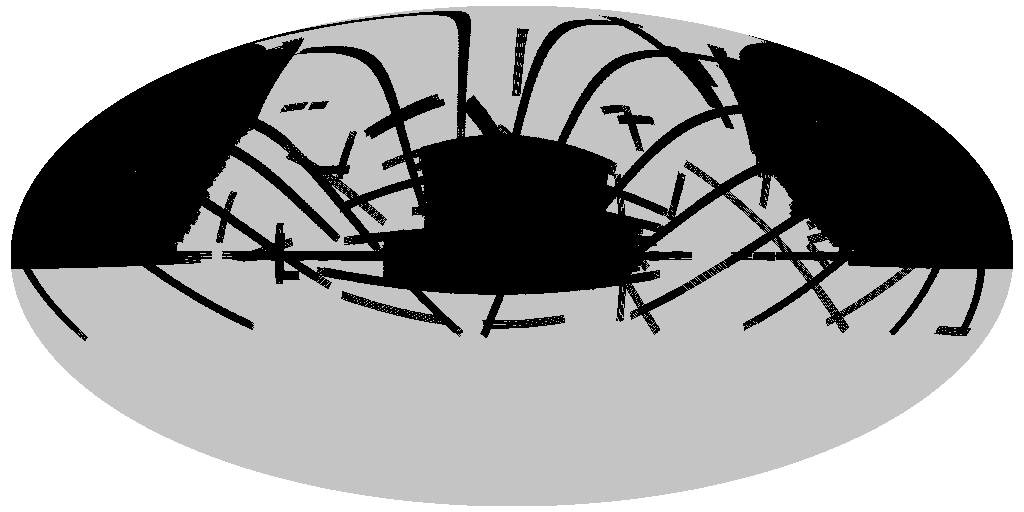}\hfill
\includegraphics[width=0.5\textwidth]{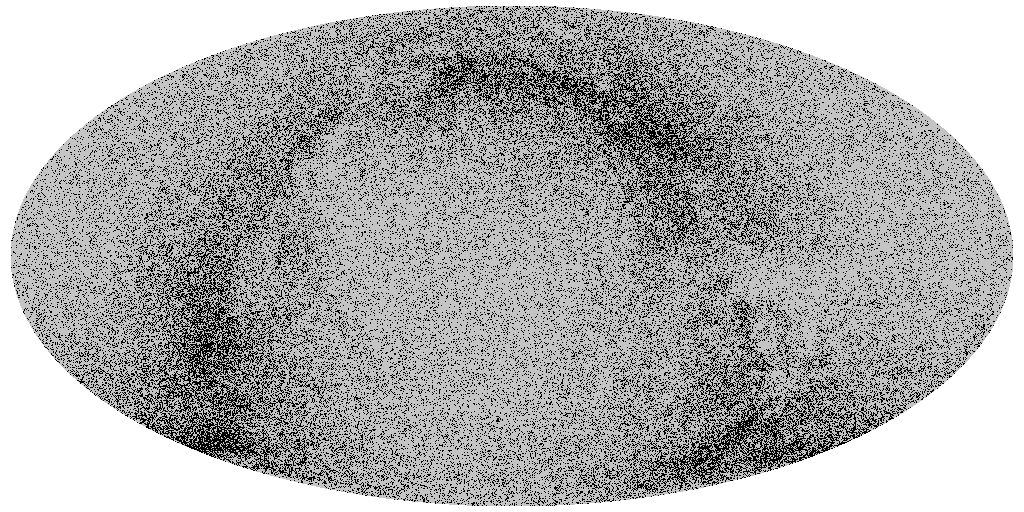}\\
\includegraphics[width=0.5\textwidth]{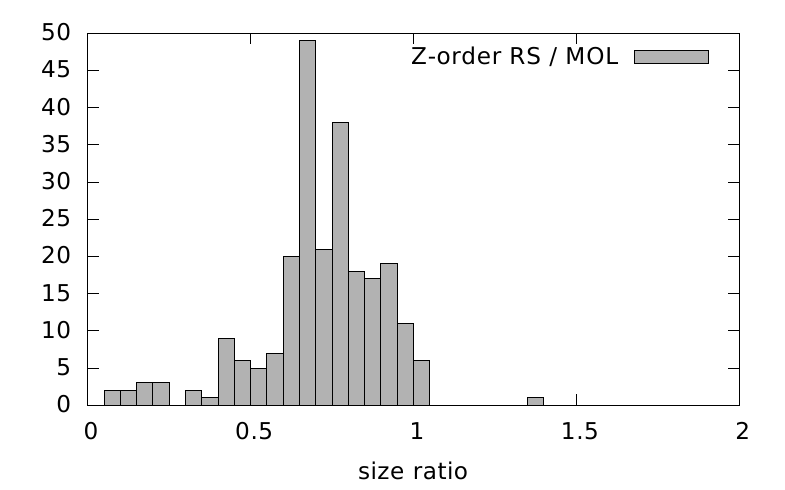}\hfill
\includegraphics[width=0.5\textwidth]{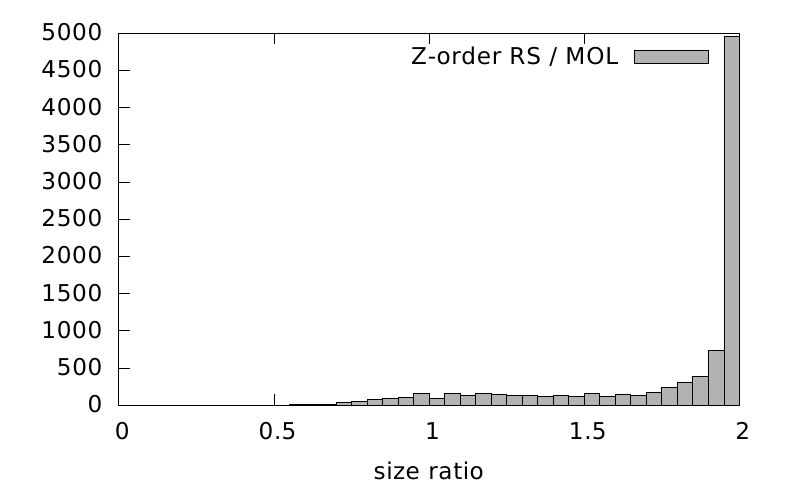}\\
\includegraphics[width=0.5\textwidth]{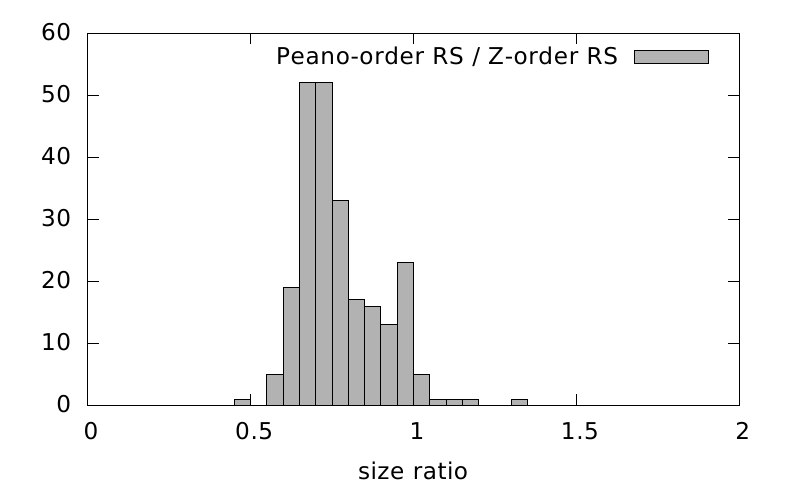}\hfill
\includegraphics[width=0.5\textwidth]{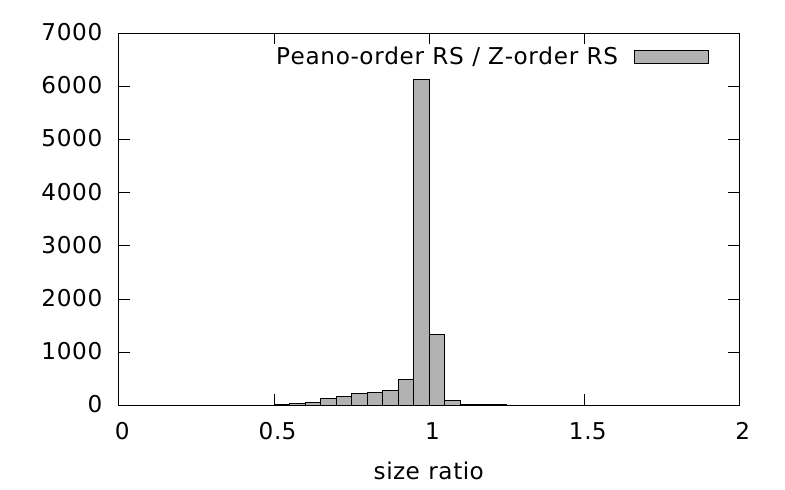}\\
\includegraphics[width=0.5\textwidth]{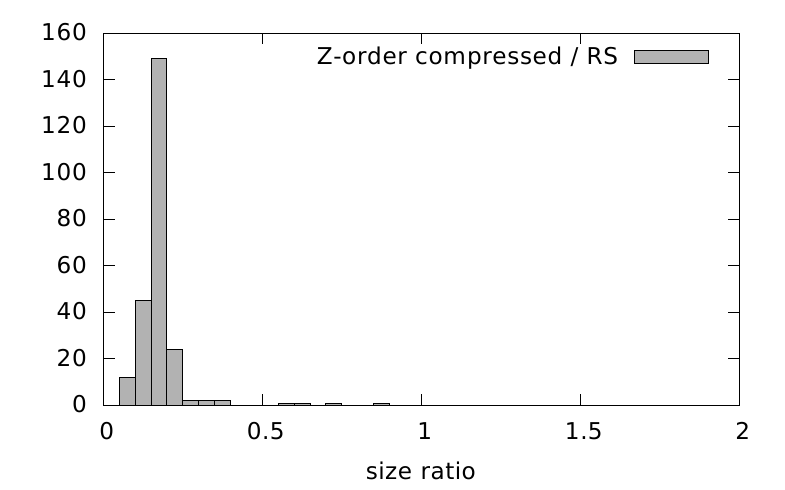}\hfill
\includegraphics[width=0.5\textwidth]{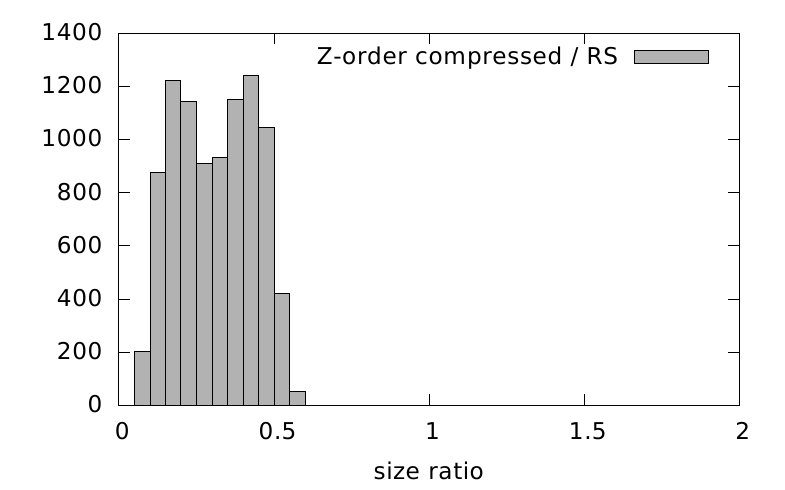}
\caption{Size comparisons for survey-like (left column) and catalogue-like (right column) spherical coverages.
The first row shows a typical representative for the respective coverage. The second, third, and fourth rows show histograms of size ratios for range sets in Z-order pixel numbering compared to multi-order lists, Peano-Hilbert RS compared to Z-order RS, and compressed to uncompressed Z-order RS, respectively.}
\label{figs:memory}
\end{figure*}

As already mentioned in Sect.~\ref{sect:inmemory}, the best choice for a small representation of a shape depends on its compactness. Regular, non-convoluted shapes are best represented as RSs, whereas multi-order lists have the size advantage in the case of strongly fragmented shapes. The available test data set contains representatives of both types, with the fragmented shapes dominating quite strongly.

To demonstrate the performance of the various representations more clearly, we split the available data sets into two groups, using the phenomenological quantity $f$ (for fragmentation)
\begin{equation}
f:=\frac{n_{\text{ranges},Z}}{n_{\text{pix,max}}}\text{.}
\end{equation}
Here, $n_{\text{ranges},Z}$ denotes the number of pixel ranges in a Z-order RS representation of the shape, and $n_{\text{pix,max}}$ is the number of individual pixels at the maximum resolution used for the shape's description. Here,
$f$ ranges from 0 (very compact shape) to 1 (extremely fragmented); we use a threshold of $f=0.1$ to discern between compact or ``survey-like'' and fragmented or ``catalogue-like'' shapes.

Furthermore, we exclude all ``small'' shapes from our size comparisons, whose multi-order list representation has fewer than 100 entries. This is done because size ratios computed for such shapes tend to produce extreme values, although their total resource consumption is close to negligible compared to the other shapes in the test data set, whose multi-order lists have up to several million entries.

The results of the memory benchmarks are shown in Fig.~\ref{figs:memory}. The results
of the size comparison of range sets to their corresponding multi-order lists are in good agreement with the estimates given in Sect.~\ref{sect:rs}. For survey-like shapes, the range set representation tends to be consistently smaller than the multi-order list, while for catalogue-like shapes it is larger by a factor of almost 2.

The improved locality of the Peano-Hilbert ordering clearly has an effect on the RSs for compact shapes, reducing their size by roughly 25\% on average. Again, the same is not true for fragmented shapes, because physically separated individual pixels do not benefit from the change in the numbering scheme overall.

Using binary interpolative coding to compress the Z-order range sets is effective for both kinds of shapes, but here as well the benefit for compact shapes is greater. In any case the results demonstrate that RSs compressed by interpolative coding are substantially smaller than uncompressed multi-order lists in almost all cases.

\section {Summary and outlook}
\label{sect:summary}
We have presented and evaluated different approaches to representing coverage information on 2D grids, while investigating alternative pixel numbering schemes and data structures. The most promising candidates for quick data processing and long-term storage were selected and implemented as part of the C++ \Healpix\ package. Correctness and performance were evaluated using a comprehensive set of VO data.

The new functionality has also been integrated in to the \Healpix\ Java library to allow easier use by software for the Virtual Observatory community, which is largely implemented in that language. Both implementations are available under the terms of the GNU General Public License v2 from the \Healpix\ Subversion repository at \url{http://sourceforge.net/projects/healpix}, and they will be part of the next official release of the \Healpix\ package.

Numerical experiments performed with the code indicate that the recommendations given in the MOC standard \citep{boch-etal-2014} concerning the data format for shapes on the sphere and Boolean operations between them can be improved upon both in terms of required memory and CPU time, although not generally in both simultaneously. The compression scheme presented in this paper is generally more space-efficient than the multi-order list described in the standard, and the RS representation allows quicker Boolean operations, at the cost of requiring potentially more memory than the multi-order list.

For applications that use pixel numbers for indexing objects in a data base, we recommend a Peano-Hilbert-based hierarchical ordering scheme instead of Z ordering, because the superior locality properties of the former scheme lead to better clustering of database accesses for queries of compact shapes.

The compact description of arbitrary shapes on discrete grids using the pixel-numbering schemes and formats presented in this paper are by no means restricted to two spatial dimensions. Extension to higher dimensions is straightforward. The only ingredient requiring nontrivial changes are the hierarchical numbering schemes presented in Section \ref{sect:numbering}, but both Z curves and Peano-Hilbert curves are available in higher dimensions, so that this part does not present a problem. All other parts of the paper are based on inherently 1D pixel indices and are not affected by dimensionality changes except for a few constants.

\begin{acknowledgements}
MR is supported by the German Aeronautics Center and Space Agency (DLR), under
programme 50-OP-0901, funded by the Federal Ministry of Economics and
Technology. We are grateful to Pierre Fernique and Thomas Boch for providing us with a very extensive test data set, and we thank No\"emi Zimdahl for the illustration in Figure \ref{figs:demo}.
\end{acknowledgements}

\bibliographystyle{aa}
\bibliography{planck}

\end{document}